# Simulated Cytoskeletal Collapse via Tau Degradation

Austin Sendek[1,2,3], Henry R. Fuller[2,3], N. Robert Hayre[2,3], Rajiv R. P. Singh[2,3], Daniel L. Cox[2,3]*

1 Department of Applied Physics, Stanford University, Palo Alto, California, United States of America, 2 Department of Physics, University of California Davis, Davis, California, United States of America, 3 Institute for Complex Adaptive Matter, University of California Davis, Davis, California, United States of America

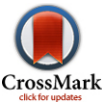

## Abstract

We present a coarse-grained two dimensional mechanical model for the microtubule-tau bundles in neuronal axons in which we remove taus, as can happen in various neurodegenerative conditions such as Alzheimers disease, tauopathies, and chronic traumatic encephalopathy. Our simplified model includes (i) taus modeled as entropic springs between microtubules, (ii) removal of taus from the bundles due to phosphorylation, and (iii) a possible depletion force between microtubules due to these dissociated phosphorylated taus. We equilibrate upon tau removal using steepest descent relaxation. In the absence of the depletion force, the transverse rigidity to radial compression of the bundles falls to zero at about 60% tau occupancy, in agreement with standard percolation theory results. However, with the attractive depletion force, spring removal leads to a first order collapse of the bundles over a wide range of tau occupancies for physiologically realizable conditions. While our simplest calculations assume a constant concentration of microtubule intercalants to mediate the depletion force, including a dependence that is linear in the detached taus yields the same collapse. Applying percolation theory to removal of taus at microtubule tips, which are likely to be the protective sites against dynamic instability, we argue that the microtubule instability can only obtain at low tau occupancy, from 0.06–0.30 depending upon the tau coordination at the microtubule tips. Hence, the collapse we discover is likely to be more robust over a wide range of tau occupancies than the dynamic instability. We suggest *in vitro* tests of our predicted collapse.

Data Availability: The authors confirm that all data underlying the findings are fully available without restriction. We have posted all the data for the microtubule simulations at http://dx.doi.org/10.6084/m9.figshare.1044310 http://dx.doi.org/10.6084/m9.figshare.1023176 http://dx.doi.org/10.6084/m9.figshare.1023087

Funding: This work was supported by United States National Science Foundation Grant DMR-1207624 (A.S., H.F., N.R.H., R.R.P.S., D.L.C.). It was also supported by International Institute for Complex Adaptive Matter, United States National Science Foundation Grant DMR-0844115 (computing cluster) (N.R.H, D.L.C.). The funders had no role in study design, data collection and analysis, decision to publish, or preparation of the manuscript.

Competing Interests: The authors have declared that no competing interests exist.

* Email: dlcox@ucdavis.edu

## Introduction

Neurodegenerative conditions such as Alzheimer's disease (AD), Parkinson's, and those resulting from chronic traumatic encephalopathy (CTE) damage arising in contact sports, represent a massive public health threat that annually impacts tens of millions worldwide. Finding routes to prevention or treatment prior to irreversible changes in the affected cells remain important goals. Part of the difficulty with this task is the complexity of the conditions themselves. For example, in Alzheimer's disease, while it is generally agreed, per the "amyloid cascade hypothesis" (see [1] and [2]) that the initial trigger of AD is the production and subsequent aggregation of A-beta protein (ABP) into oligomers (ABO), the way in which the ABOs trigger cell damage and eventual death remains unclear.

Neuro-fibrillary tangles (NFTs), consisting largely of hyperphosphorylated tau protein aggregates, have been observed in post-mortem tissues of AD victims. These NFTs are correlated with regions of A-beta aggregates and neuronal damage or death [3] confirms that ABOs trigger processes which lead to modifications of the tau protein. NFTs and tau associated damage also arise in CTE [4]. Tau proteins play a critical role in the microtubule bundles (MTBs) of the axon cytoskeleton, by promoting the assembly of axonal microtubules and inhibiting dynamic instability

(DI) [5], and by crosslinking the microtubules in the bundles. The integrity of the MTBs is critical for neuronal function: they provide mechanical stability to axons for decades of human life, and they serve as the highways for fast and slow axonal transport of neurotransmitters and organelles to and from the synaptic regions of the neuron. Clearly, at the cellular level, a significant marker of irreversible damage in conditions affecting tau proteins would be loss of integrity of the MTBs, and thus a point of no return for intervention.

Studies of cultured neurons exposed to high concentrations of ABOs have confirmed that the ABOs can be internalized, and once internalized these trigger production of proteinases which can cut taus, and kinases which can phosphorylate taus [6]. Both are important, since full length taus are needed to assure the proper spacing of microtubules, and hyperphosphorylated taus (HPTs) may tend to dissociate from the negatively charged microtubule surfaces.

If tau detachment from microtubule binding sites exceeds 50%, this should lead to collapse of MTBs due to the dynamic instability of individual microtubules. In experiments on cultured hippocampal neurons exposed to massive concentrations of ABOs (1000–10000 times physiological ABP concentrations *in vivo*) MT loss is indeed observed [7], with the lateral density decreasing by a factor of 25 compared to unexposed cells. However, these high

ABO concentrations may obscure more subtle changes in the MTBs. Specifically, since the taus provide transverse mechanical stability, MTB transverse rigidity may be lost when a majority of the taus are still intact: random removal of connecting springs in a lattice of otherwise noninteracting cylinders leads to a loss of transverse mechanical stiffness at the rigidity percolation threshold near 60% of spring occupancy as we discuss below. Moreover, as the HPTs no longer favorably associate with the surface of microtubules, they can lower the free energy by gaining translational entropy through reduced excluded volume between microtubules, yielding a fluctuation mediated attraction between microtubules. This will, in effect, add an inward radial pressure to the MTBs as taus are removed.

In this paper, we develop a two dimensional mechanical model for the MTBs and consider the percolation transition as we remove taus between the microtubules (Fig. 1). Our method confirms that in the absence of the depletion force, the transverse rigidity is lost when the concentration of tau binding is reduced to nearly 60% of the original value. We denote this ratio of bound tau concentration to the maximum value as the tau occupancy $p$. Adding the depletion force via a simple excluded volume model leads to a first order collapse that depends upon the ratio of the tau spring force to the concentration of HPT. This collapse arises because the steady inward pressure of the nearest neighbor MT mediated depletion force contracts the microtubule bundles to a point where next-nearest-neighbor microtubules can experience attraction. For physiologically attainable values of tau spring force and HPT concentrations, the collapse transition can easily arise above 40% bound tau occupancy, while we argue that the dynamic instability is likely to occur below this value. Hence, with or without the depletion force we believe there will be irreversible mechanical damage to the MTBs at high bound tau concentrations, well before the dynamical instability is relevant. In formulating treatment strategies for tau related diseases, these conclusions could be very important.

## Results

### Overview of Model

Consistent with experiment, the MTB is modeled initially as a simple hexagonal lattice of one micron long rigid microtubules of radius $R = 12.5 nm$ (blue disks in Figure 1) linked by tau springs (red lines). There are approximately $\lambda = 50$ tau springs per micron in a microtubule pair with spring constant $k_{eff}$, and the mean microtubule center to center distance is taken to be 45 nm [8]. The springs are modeled harmonically in this initial simulation where compression is small on a per spring basis. We have simulated five MTB sizes of varying initial radii $R_{mtb}(1)$ at full tau occupancy $p = 1$: $R_{mtb}(1) = $ 100, 150, 200, 250, 300 nm with, respectively, 19, 37, 73, 117, and 163 microtubules. The radius is measured as the maximum center-to-center MT separation from the central MT to a boundary one for the hexagonal shaped bundles. Because each 50 nm increase in radius adds one hexagonal shell of microtubules, we describe bundle sizes using "$n$-shell" where $n = $ 2, 3, 4, 5, 6 for $R_{mtb}(1)$. We neglect surface tension at the boundary due to the combined action of the axon membrane and actin cytoskeleton, as direct measurement shows this to be small for axons [9]. We employ the steepest descent relaxation algorithm (equivalent to highly damped molecular dynamics), where microtubules are iteratively moved a distance proportional to the net force experienced until the positions are converged to within a fractional tolerance factor [10]. We present details of the relaxation algorithm in the Supplemental Information. The ratio of initial resting spring length $\ell_0$ to lattice spacing $a$ is denoted $\eta$, which we vary in the case of no depletion force. We note that we are not including neurofilaments in our model, which are not critical for stability of the MTBs (though they are for axons) [11].

The taus are intrinsically disordered proteins with a single microtubule binding domain. Hence, they must be at least in dimer form to link two microtubules [12]. The N-terminus of the tau is predominantly negative, while the middle (M) region preceding the microtubule binding domain is predominantly positive. As argued in Ref. [12], the tau can dimerize in this region through salt bridges between the N/M regions, which take up about 200 residues, giving a total peptide length to this domain of $L_{NM} = 200 a_R \approx 80$ nm where $a_R$ is the length of a residue, about 0.4 nm. The simplest model for the spring constant, in the low extension/compression regime, is of a paired set (one per monomer) of $(n_s - 1)$ entropic springs of length $\ell_s = L_{NM}/(n_s - 1)$ between $n_s$ salt bridges where we assume the unstructured entropic tau peptides have a short persistence length $\xi_p \simeq 1$ nm. The result is (c.f., this result in Phillips et al.) [13]

$$k_{eff} = \frac{2(k_B T)}{2(n_s - 1)\ell_s \xi_p} = \frac{k_B T}{L_{NM} \xi_p}. \qquad (1)$$

We thus estimate $k_{eff} \approx 0.05$ pN/nm. Proteins with well defined secondary structure of course have much higher effective spring constants under compression, potentially reaching more than 10 pN/nm, as can be inferred from works on green flourescent protein [14] and fibrinogen [15] for example.

Figure 2 shows the excluded volume model for the depletion force, derived from cylindrical geometry following the spherical example of Phillips et al. [16]. Once taus are phosphorylated, they no longer associate with the microtubules with the annulus between $R$, $R + r$. To leading order in the volume difference, the effective interaction potential $V_D(Z)/L$ per unit length $L$ between a pair of microtubules with center-to-center distance $Z$ and $Z/(2(R + r)) = \zeta$ is given by

$$\frac{V_D(Z)}{L} = \frac{P_O}{2} Z^2 \left( \sqrt{\frac{1}{\zeta^2} - 1} + \frac{1}{\zeta^2} (\sin^{(-1)}(\zeta) - \frac{\pi}{2}) \right) \theta(1 - \zeta). \qquad (2)$$

Here $P_O$ is the osmotic pressure of the nonbinding hyperphosphorylated tau dimers given by

$$P_O = k_B T [\tau_{2p}] \qquad (3)$$

where $k_B$ is Boltzmann's constant, $T$ the temperature, and $[\tau_{2p}]$ the concentration of hyperphosphorylated tau dimers. The corresponding depletion force is taken as the negative gradient of $V_D$ with respect to $Z$. This gives a depletion force per unit length $F_D(Z)/L$ of

$$\frac{F_D(Z)}{L} = -P_O Z \left( \sqrt{\frac{1}{\zeta^2} - 1} \right) \theta(1 - \zeta).$$

In our simulations, of course the tau dimers are not spherical objects, and their precise radius of gyration is not known, although we anticipate the dimers to be at least 20 nm long (the wall-to-wall distance between microtubules) and 5–10 nm thick. For simplicity we have taken $r = R = 12.5$ nm. We find, in the region of collapse($Z \approx 35nm$), that the magnitude of the depletion force per unit length divided by the osmotic pressure $\phi_D(Z)$ is weakly dependent upon $r$ and given approximately by

$$\phi_D(Z) = \frac{F_D(Z)}{LP_O} \approx -2.0 - 5.5r \ nm. \qquad (4)$$

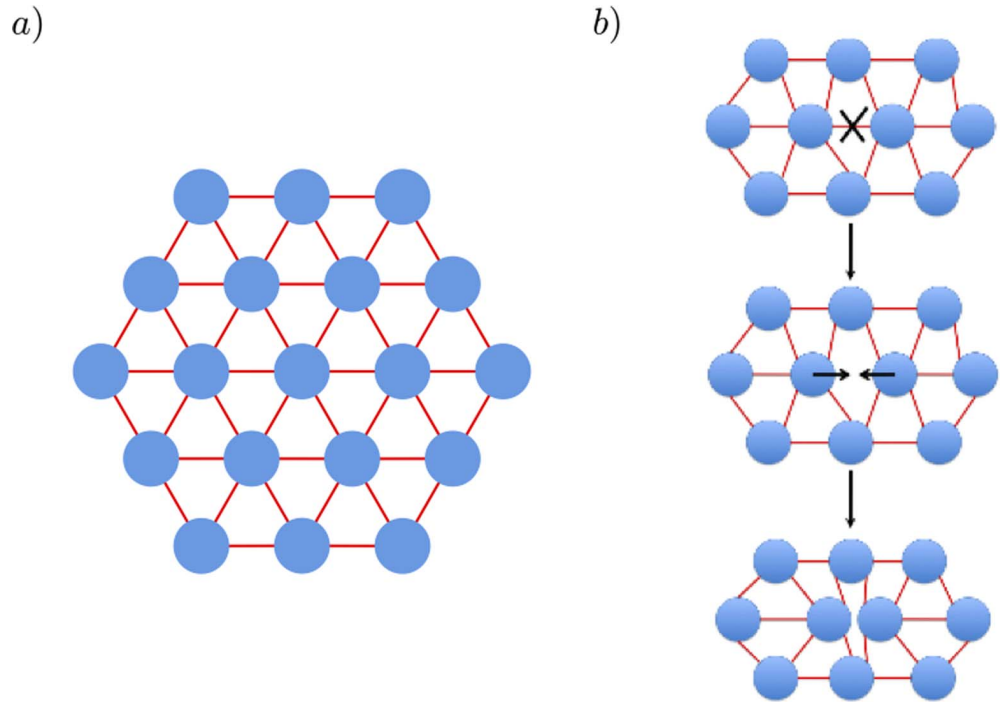

**Figure 1. Two Dimensional Microtubule Bundle (MTB) Model.** Microtubules are treated as rigid cylinders of diameter 25 nm (blue disks) with center-to-center distance of 45 nm (a). Taus are treated as springs, with 50 per micron length of a microtubule pair. To model tau depletion, taus are removed at random and the system is re-equilibrated with the steepest descents method described in the text (b).
doi:10.1371/journal.pone.0104965.g001

$\phi_D$ varies only from $-56$ nm to $-85$ nm as $r$ varies from 10–15 nm. Hence, the main determinant of the depletion force is the magnitude of the osmotic pressure.

We note that taking the radius of gyration of the tau dimers to be 10–15 nm makes them larger than the observed (calculated) value of 6.5(6.9) nm for tau monomers [17]. However, these monomers are in a compressed conformation in which the positive and negative charges comingle. The relatively extended conformation necessary for the dimer must be larger.

The upper limit on the osmotic pressure associated with hyperphosphorylated tau dimers is clearly the bound concentration of dimers, $[\tau_{2b}]$, in the nondegraded axon. Using the volume between microtubules, we estimate $[\tau_{2b}] = 320$ micromolar, which gives an osmotic pressure of 800 Pa. We note that the density of bound taus in the MT bundles is significantly larger than the $\simeq 2$-4 micromolar estimated concentrations of free tau monomers, as has been observed before [18].

It is convenient to characterize our model bundles with depletion force in terms of dimensionless ratio $\rho$ of the spring force constant per unit length compared to the osmotic pressure, which is given by

$$\rho = \frac{\lambda k_{eff}}{P_O}. \qquad (5)$$

For $\lambda = 50/$micron, $k_{eff} = 0.05$ pN/nm, and $P_O = 800$ Pa, $\rho = 3.125$.

## 0.1 Test Case: simulation of taxol stabilized microtubules with PEO intercalants

We note that the depletion force has been observed in *in vitro* experiments with taxol stabilized microtubule bundles intercalated with PEO polymers (and no taus) [19]. In this case the only additional interaction between the micotubules is via screened Coulomb coupling, which is estimated by the force per unit microtubule surface area $P_{ES}$

$$P_{ES}(Z) = 0.078 \exp(-(Z-2R)/\Lambda_D) \frac{nN}{nm^2} \qquad (6)$$

with the Debye length $\Lambda_D = 1.47$ nm. We add this force to our depletion force to test our steepest descents equilibration as shown in Fig. 3. Over a factor of 100 variation in $P_O$, the agreement between theory and the data in Fig. 11 of Needleman *et al.* [19] is good. Note that given the large rest length of the taus, the interaction $P_{ES}$ is irrelevant prior to any collapse and so we neglect it in our model for neuronal microtubules.

## 0.2 Model MT repulsion for axonal MTBs

In simulations with taus for axonal MTBs, we mimic the combined screened electrostatic repulsion/steric repulsion between microtubules by a Morse potential of the form

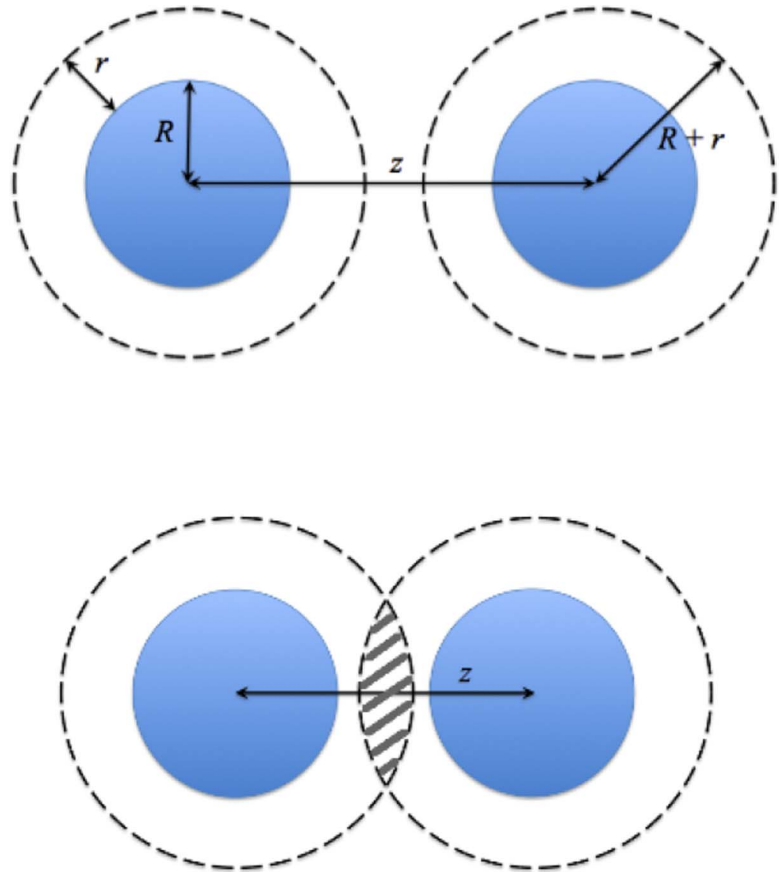

**Figure 2. Depletion mediated interaction between microtubules.** For intercalants of radius r (presumed here to be hyperphosphorylated taus) an annulus between $R, R+r$ is depleted of intercalants. The loss of excluded volume as the microtubules approach allows increased translational entropy outside the bundle for the intercalants.
doi:10.1371/journal.pone.0104965.g002

$$V_M(Z) = D_e[1 - \exp[-a(Z-2R)]]^2 \qquad (7)$$

where we take $D_e = 10^{-4} J$ and $a = 1$ nm$^{-1}$. While this potential does provide a small minimum near the microtubule surfaces, its main effect is to inhibit collapse beyond the region where microtubules are in contact, which is sufficient for our purposes.

## Test case: Rigidity Percolation with no Depletion Force

In the depletion force free case, we have computed the second order elastic constant $C$ (Fig. 4) which measures the stiffness of the microtubules against transverse compression for a variety of different ratios $\eta$ of the initial microtubule-to-microtubule separation to spring resting length. In all cases, the scaled curve for $C(p, \eta)/C(1, 1)$ is identical, since unlike the work of Ref. [20] we do not hold our springs in a fixed frame under tension but rather allow them to relax, so the result is essentially equivalent to the case where bond length equals rest length in Ref. [20]. This gives a transverse rigidity collapse (zero transverse elastic constant) at a critical tau occupancy $p_c \approx 0.63$ in agreement with Ref. [20]. Note that we do not see a sharp threshold due to finite size effects; we have confirmed that the rounding of the transition is reduced as a function of increasing system size. We stress here that by vanishing elastic constant, we mean that the MTB will initially lose rigidity against radial compression, until the MTs can come into direct steric contact, at which point the elastic constant will jump to a value suitable for close packed MTs. Hence, even without inclusion of the depletion force, we anticipate a significant mechanical degradation (transverse rigidity loss) at a high $p$ value which exceeds the range ($p \leq 0.3$) argued below to be relevant for the dynamic instability mechanism.

## MTB Collapse with Depletion Force

With the depletion force present, we find a first order volume collapse as $p$ is reduced, as shown in Fig. 5, for one value of the dimensionless parameter $\rho$. This figure shows a typical plot of the normalized transverse radius of gyration of the microtubule bundle $R_{MTB}(p)/R_{MTB}(1)$ decreasing slowly with decreasing $p$ until the critical percolation threshold $p_c$ is reached, at which point $R_{MTB}$ drops precipitously to the hexagonal closed packed steric limit $R_{MTB}(p_c) = 2nR = 25n$ nm, where as defined above, $n$ is the number of hexagonal shells. The parameter $\rho$ controls the position of the collapse, as shown in Fig. 6(a), where we show curves for $p > p_c$ such that the left endpoint is at $p_c$. This collapse is independent of the size of our starting bundle showing that it is not an artifact of small bundles in our simulation as shown in Fig. 6(b) where we plot the results for a single $\rho$ value several bundle sizes. Each curve in Fig. 6 is averaged over five runs and is truncated at $p = p_c$; the horizontal bar located at the collapse region represents (by its width) the range of $p_c$ values obtained. Fig. 7 shows that holding $\rho$ fixed while co-varying $P_O, k_{eff}$ yields the same collapse curves for $p > p_c$. Note that below the collapse the compressional rigidity will increase, as it is no longer dominated by the taus but by the MTs themselves (which become close packed below the collapse). We illustrate the collapse in the movie (MovieS1.mp4) provided in the supplemental material, which depicts the $n=2$ bundle as taus are removed.

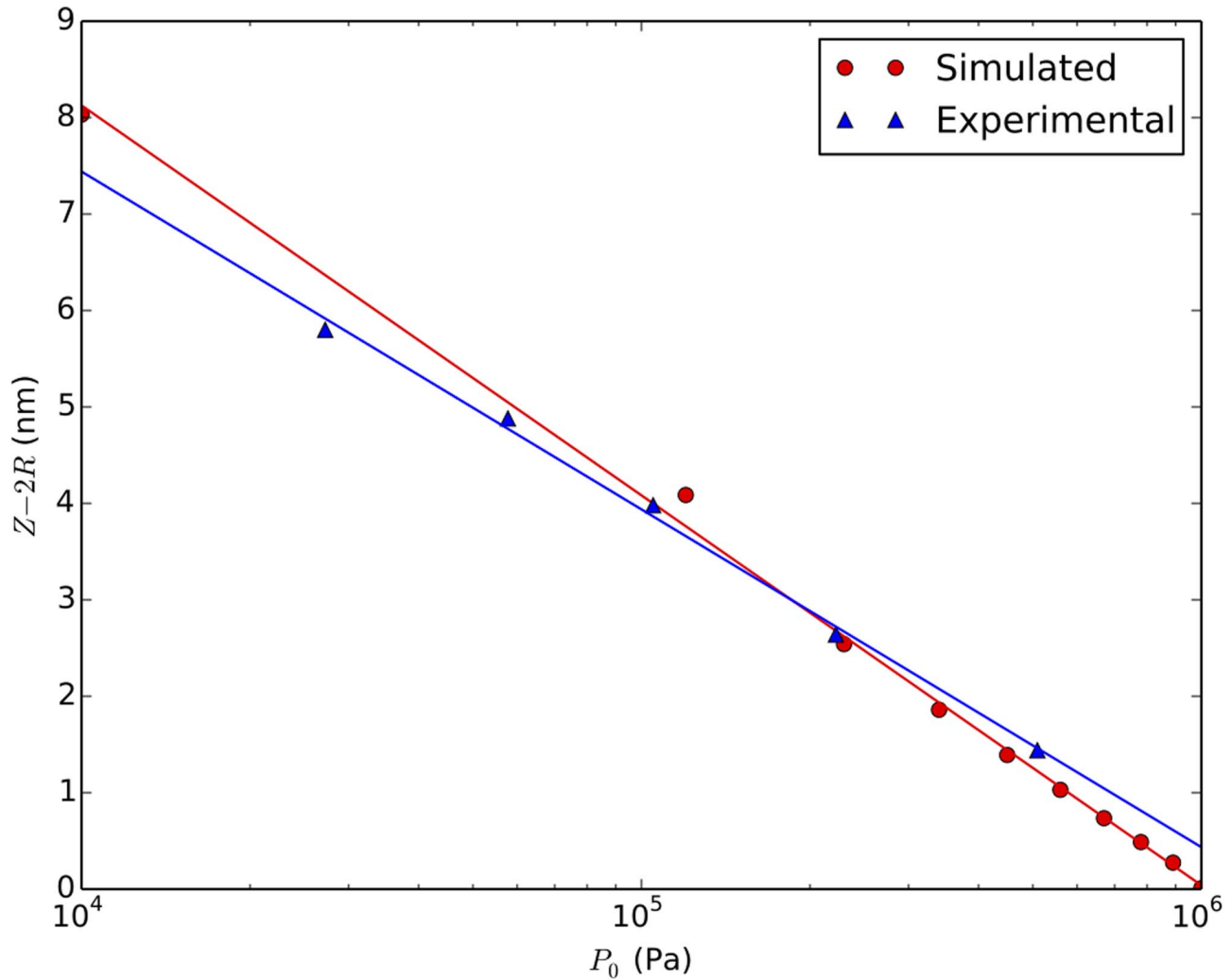

**Figure 3. Simulation of equilibrated force vs. distance for taxol stablized microtubule bundles with PEO intercalant per Ref. [19].**
The red circles are our data for osmotic pressures given in Ref. [19], the blue triangles are the data in that reference.
doi:10.1371/journal.pone.0104965.g003

In the majority of of our simulations we have taken the osmotic pressure constant at all $p$ values, but of course if the depletion force is due to the phosphorylated tau dimers, we expect $P_O$ to be proportional to $1-p$. The lavender curve in Fig. 6(a) shows the result for which the osmotic pressure is taken to be proportional to $1-p$, reaching the same value as for the yellow curve at the collapse threshold $p=p_c$. Clearly to within the range of $p_c$ values obtained over our different random number seeds, we obtain the same $p_c$ for constant and varying $P_O$ models. The lavender curve varies more substantially in radius, because at $p\rightarrow1$, there is a different equilibrated radius suitable for zero depletion force.

## Heuristic Mean Field Theory (MFT) of MTB Collapse

The origin of the collapse becomes clear if we consider a heuristic mean field theory (MFT) for the microtubule bundles. The assumptions which go into this are as follows: (i) as we remove taus we replace $k_{eff}$ with $pk_{eff}$. (ii) The average symmetry around a given microtubule remains hexagonal, and the nearest neighbor and next nearest neighbor coordination numbers are 6. (iii) There is a hard wall when the microtubule center-to-center separation reaches $2R$. (iv) Initially, the next nearest neighbor microtubules are at distances $R_{NNN} \geq 2(R+r)$, but as the system contracts, they reach within the depletion force active zone.

Because of the latter assumption, the potential energy acquires an extra "kick" when the next-nearest neighbor MTs enter the depletion zone radius. A simple model energy per unit length $\epsilon_{MFT}$ which captures this is

$$\epsilon_{MFT} = \frac{E_{MFT}}{L} = \frac{1}{2}p\lambda k_{eff}(Z-\ell_0)^2 + V_D(Z)/L + V_D(\sqrt{3}Z)/L \quad (8)$$

where $Z$ is the mean MT center-to-center separation and $L$ is the MT length. Here $E_{MFT}$ is the total energy of the MTB within the mean field theory. The third term accounts for the next nearest neighbor interactions. We supplement this by an infinite wall at $Z \leq 2R$.

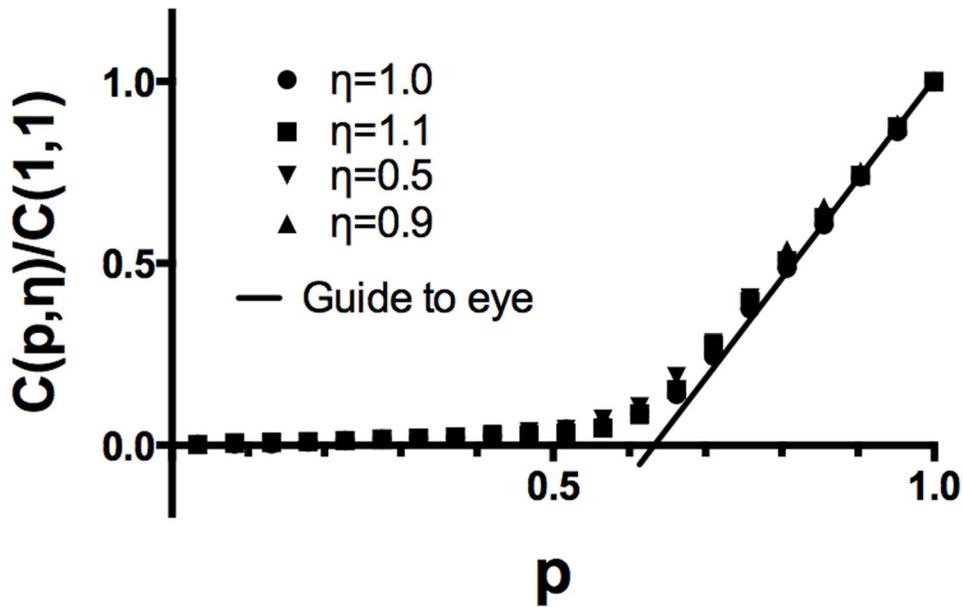

**Figure 4. Transverse rigidity transition in the absence of the depletion force.** As described in the text, we compute the 2nd order elastic constant for radial compression as a function of the occupied spring fraction $p$ and (b) scale to the $p=1, \eta=1.0$ limit. In contrast to Ref. [20], all $\eta$ values yield the same results when scaled, because their lattice boundary is fixed while ours relax to equilibrium. In all cases, the transverse rigidity percolation threshold is $p_c \approx 0.6$, in good agreement with the $\eta=1.0$ case of Ref. [20]
doi:10.1371/journal.pone.0104965.g004

The key idea of the MFT is illustrated in Fig. 8. There, we show several curves for $v_{MFT} = \epsilon_{MFT}/(2P_O(R+r)^2)$ vs. $\zeta = Z/2(R+r)$ as $p$ is varied. For $p > p_c$, e.g., $p=1$, there is a minimum at an equilibrated value $Z \approx 1.6(R+r)$ due to the balance of depletion attraction and spring repulsion. As shown in Fig 7 for $2(R+r)$ $=50$ nm and $\rho=3.125$, the enforced hard wall minimum at $Z=2R$ becomes degenerate with the large $Z$ minimum as $p$ is reduced to the critical value $p_c = 0.815$. Below the critical value for $p$, the hard wall minimum is lower in energy. Hence, the transition at $p_c$ is a first order jump. In Fig. 9 we show a comparison of the normalized $p_c$ values vs. $\rho$ for the MFT and full simulation; the trends are both quantitatively and qualitatively similar.

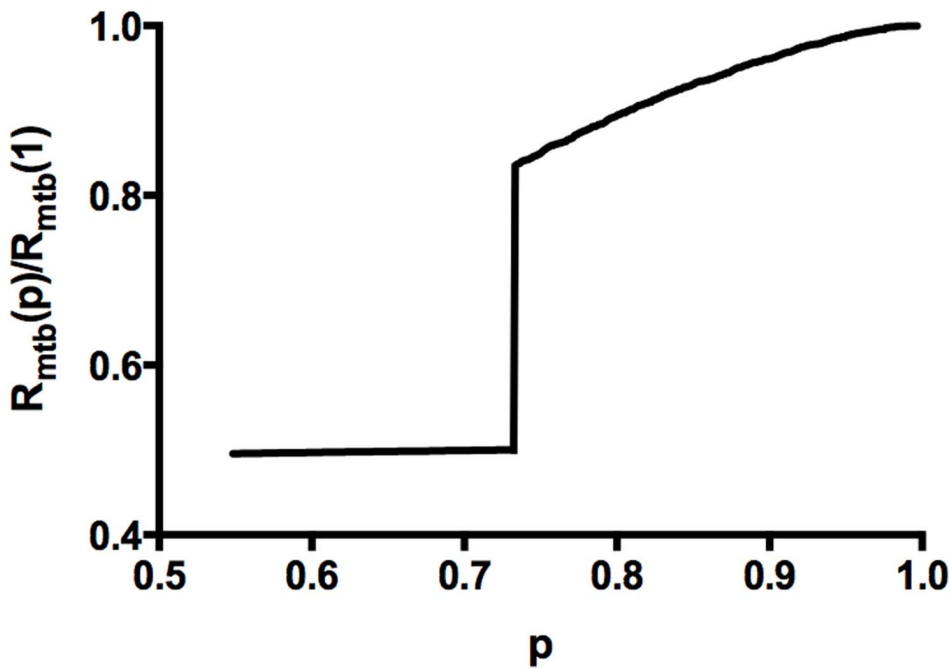

**Figure 5. First order transverse collapse in the presence of the depletion force.** This figure traces the transverse collapse as the bound tau density $p$ is reduced from 1, for the dimensionless parameter $\rho=3.125$. After a gradual reduction of $R_{MTB}$ for $p=1$, the depletion force overwhelms the entropic springs at $p_c=0.73$.
doi:10.1371/journal.pone.0104965.g005

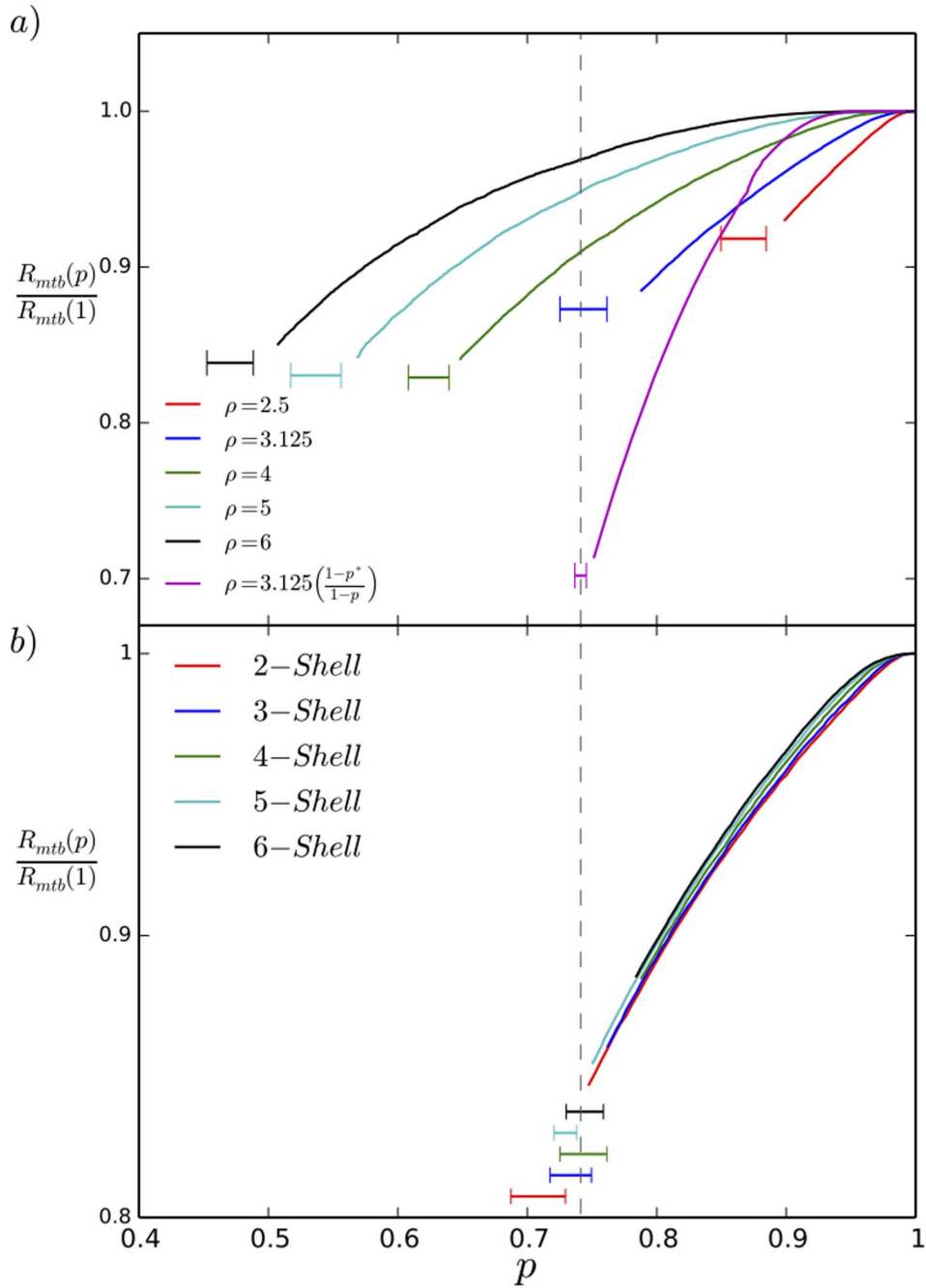

**Figure 6. Normalized microtubule bundle radius $R_{MTB}(p)/R_{MTB}(1)$ vs. tau occupancy $p$.** In all these curves we show the bundle radius above $p_c$ with the leftmost end at $p_c^\pm$. a) The MTBs undergo a first order collapse with reduced $p$, with the collapse onset decreasing with $P_O$ for fixed spring constant $k_{eff}$ = 0.05 pN/nm. The lavender curve is computed with the same $P_O$ value at collapse as the blue curve but for an osmotic pressure proportional to $(1-p)$. For $p < p_c$ (not shown) the radius $R_{MTB}$ is that of hexagaonally close packed microtubules. b) For varying initial bundle radius, measured by the number of hexagonal shells retained about the central MT, the normalized radius displays the collapse at the same location for a fixed value of $rho = \lambda k_{eff}/P_O$ =3.125, indicating the collapse is not an artifact of finite size.
doi:10.1371/journal.pone.0104965.g006

The driving feature for the collapse is the bootstrapping effect of diminished bundle radius driving increased depletion attraction which in turn pulls the next nearest neighbors in to overwhelm the resisting tau entropic springs.

Another important feature shared by the MFT and the full simulation is that a fixed $\rho$ value driven by a fixed $P_O$ will produce an MTB collapse at the same $p_c$ as does a varying $\rho(p)$ with $P_O(p) = P_O(p_c)(1-p)/(1-p_c)$. The latter model for the osmotic pressure is appropriate for the assumption that the removed taus themselves produce the depletion force.

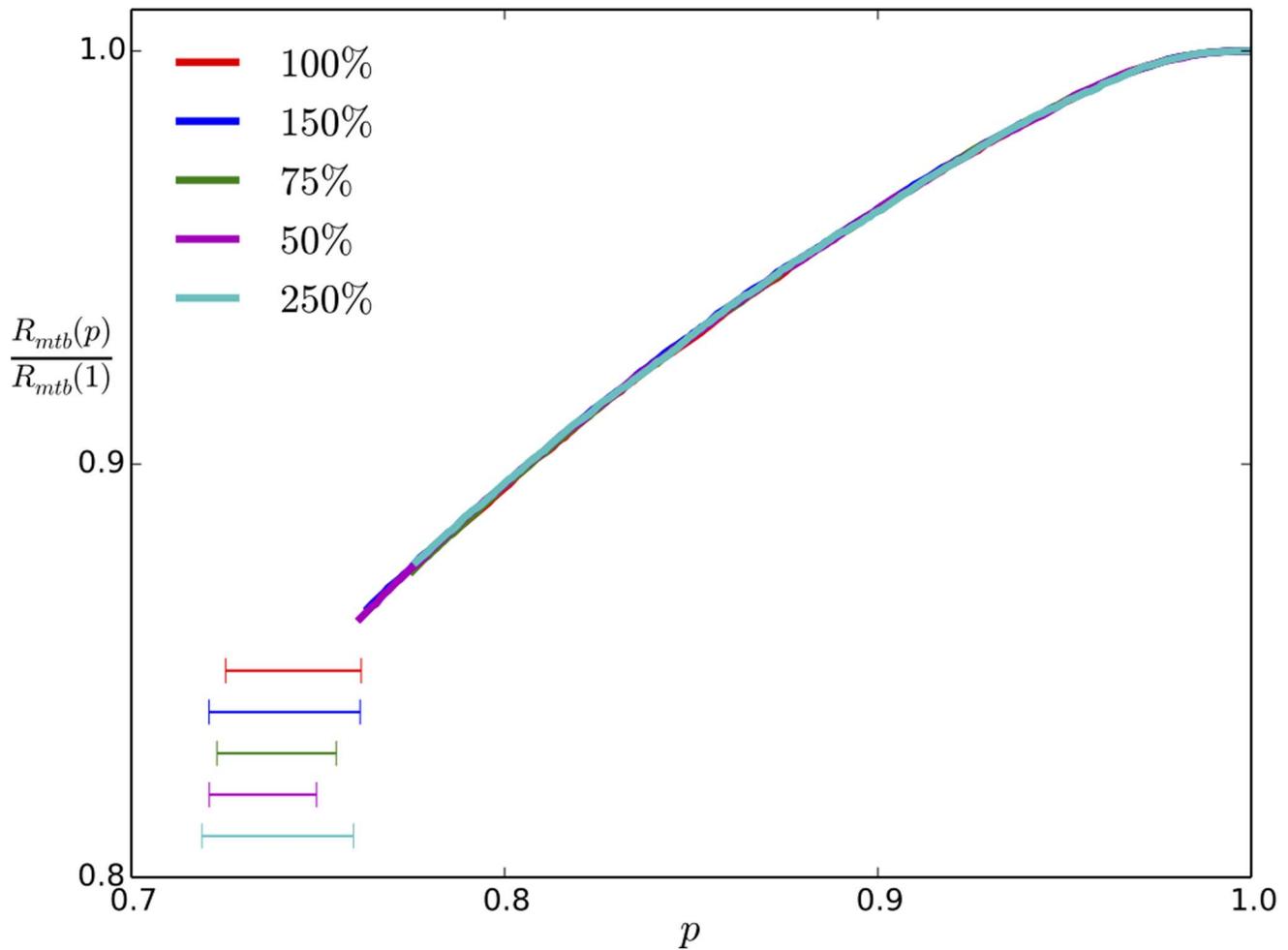

**Figure 7. Normalized microtubule bundle radius $R_{MTB}(p)/R_{MTB}(1)$ vs. tau occupancy for fixed $\rho$.** In all these curves we show the bundle radius above $p_c$ with the leftmost end at $p_c^+$. Here we carry out the simulations multiplying $k_{eff}, P_O$ independently by the percentages shown in the legend to achieve the same $\rho$ value. Clearly this gives the same $p_c$ for collapse onset.
doi:10.1371/journal.pone.0104965.g007

## Competition with Dynamic Instability: Range of possible collapse dominance

It has been presumed in previous work that the most likely route of degradation for the MTBs under tau degradation is via dynamic instability, in which the MTs stochastically vanish due to dominance of depolymerization when there are insufficient stabilizing taus. By augmenting our model with the following assumptions, we can produce a heuristic estimate for the $p$ threshold below which dynamic instability should be important:

(i) Depolymerization initiates at the ends of MTs, so the critical stabilizing molecules binding to MTs should do so at the ends.

(ii) Given ample evidence for two binding sites of tau on MTs [21], including one in the MT lumen [22], we assume that taus at the ends provide insurance against dynamic instability [23].

(iii) Evidence from studies with the small molecule eribulin suggest that binding of one molecule at the end is sufficient to inhibit dynamic instability [24].

Thus we assume that it is necessary to remove endpoint taus to generate dynamic instability. An estimate of where this will occur depends upon (i) the tau coordination at the ends, and (ii) the probability of site percolation on our lattice (if we remove a continuous backbone of microtubules through dynamic instability we lose lattice integrity). The site percolation threshold for two dimensional triangular lattices is 0.5 [25]. A conservative upper bound on the threshold for dynamic instability is thus to assume an end coordination of 1, and the probability of removing 2 end taus from a microtubule is thus $(1-p)^2$. 50% of end taus would then be removed when $(1-p_{DI})^2 = 1/2$, giving $p_{DI+} = (1-1/\sqrt{2}) = 0.29$. If we assume instead that the coordination is 6 at the ends, we obtain $(1-p_{DI-})^{12} = 1/2$ as the criterion, giving the occupancy of taus to generate DI at $p_{DI-} = (1-2^{-1/12}) = 0.056$.

In fact, our model is oversimplified by assuming that all taus bind equally to the MTs. The presence of two binding sites with different affinities (lumen binding taus have a higher affinity than surface taus) suggest that the stabilizing end taus are harder to remove with phosphorylation, so that the above estimates for the onset of DI are likely to be reduced further.

This leads us to conclude with confidence that the collapse scenario here is likely to occur over the dynamic instability for $p \geq 0.29$, and is certainly very probable for $p \geq 0.06$. With the more conservative criterion (removal of one tau from each end) then we find for $r = 15$nm that the collapse is viable over DI for $k_{eff} \leq 0.14$ pN/nm, while for $r = 10$nm the collapse is viable for $k_{eff} \leq 0.067$ pN/nm.

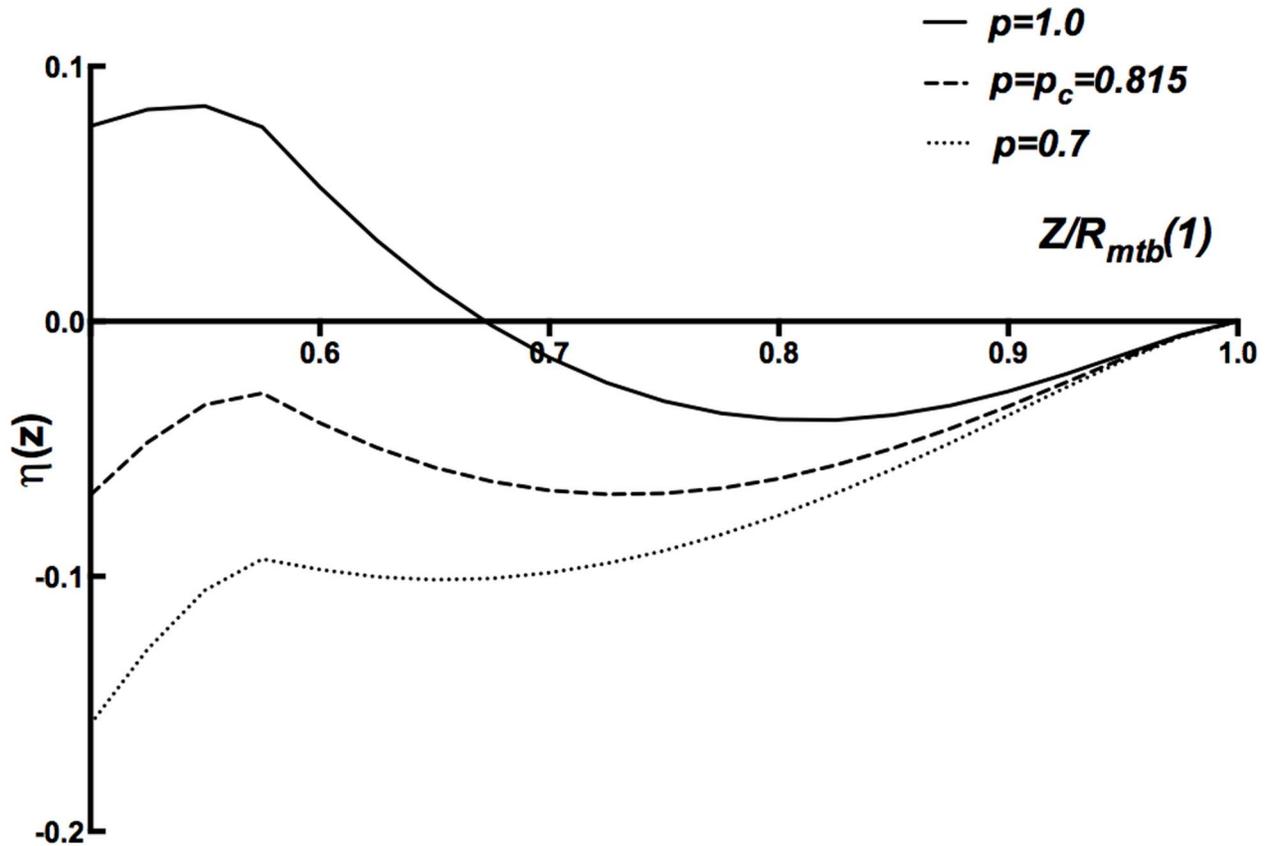

**Figure 8. Potential Energy of Mean Field Theory.** Energy per unit length $v = \epsilon_{MF}/(2P_O(R+r)^2)$ vs. scaled mean microtubule separation $\zeta = Z/(2(R+r))$ for $2(R+r) = 50$nm and $\rho = 3.125$. As $p$ is reduced, the potential at $\zeta = 0.5$, which is the separation at microtubule contact, reaches a lower value than the minimum which evolves from $\zeta = 0.82$ at $p = 1$.
doi:10.1371/journal.pone.0104965.g008

## Discussion

Having demonstrated that a MTB collapse is likely to occur for a range of reasonable physical parameters under conditions that phosphorylated taus are removed from microtubules, it is reasonable to ask whether this can be observed or may have implications correlating with existing observations.

First, as mentioned earlier, there is strong evidence of MT dynamic instability in cultured neurons exposed to massive doses of ABO over a period of hours, at 1–5 micromolar concentration [7] compared to estimated AB concentrations of .25–.5 nanomolar *in vivo* [26]. An exposure of mature cultured neurons over a period of days to more reasonable subnanomolar concentrations of AB dimers (but not monomers) does not induce microtubule damage at the resolution of the experiments but does cause beading of hyperphosporylated taus proximate to the microtubules and neuritic damage [27]. We conjecture that the high concentration ABO experiments induce massive phosphorylation of the taus, while the lower concentration exposures may not cause enough tau removal to induce our proposed MTB transverse collapse.

We propose two direct experimental tests to observe this novel collapse. First, we suggest using taxol stablized microtubule bundles *in vitro* per Ref. [19] with taus binding along the MT external surfaces, then exposing the MTBs to phosphorylating kinases. The taxols will inhibit dynamic instability, so this will be a clean test of whether the transverse collapse scenario proposed here is viable. Second, one can measure the response of the axon to ABO exposure in cultured neurons with microbead compression of the neurons. This has been done for initial ABO exposure on cultured axon free N2a cells and on HT22 mouse hippocampal cell nuclei where it has been shown that the mechanical properties of the membrane change are consistent either with increased osmotic pressure due to calcium influx or a stiffening of the membrane under 30 minutes of exposure to micromolar concentrations of ABO [28]. We would advocate monitoring of the axon directly over time with microbead compression to look for evidence of collapse in the force response curves. If the depletion force is irrelevant, then at some point the rigidity loss will lead to a change in the measured elastic constant. If there is a collapse then there will be a change in the monitored force in contact with the bead. Even at high exposures of ABO, if the collapse scenario is correct there should be an intermediate time scale in which the collapse is observable prior to the onset of MT dynamic instability. Studies of axonal compression under ABO exposure in late stages of cell life have not, to our knowledge, been carried out.

One observation of patients with AD is an atrophy of white (axonal) matter in affected regions of the brain. Functional MRI measurements of water diffusion anisotropy for patients with mild cognitive impairment gives evidence for axonal damage [29]. Additionally, it has been determined that changes in white matter volume of AD patients compared to elderly control patients can reach values of 16.4% in parahippocampal tissues and 21.3% in entorhinal tissues [30]. While cell death in late stages of AD can certainly decrease tissue volume, the cause of this atrophy in living patients has not been positively ascribed to cell death and could arise from the collapse scenario presented here, although we caution that the spatial resolution is certainly insufficient for positive attribution. Axonal dystrophy is a hallmark of later stage Alzheimer's disease [31].

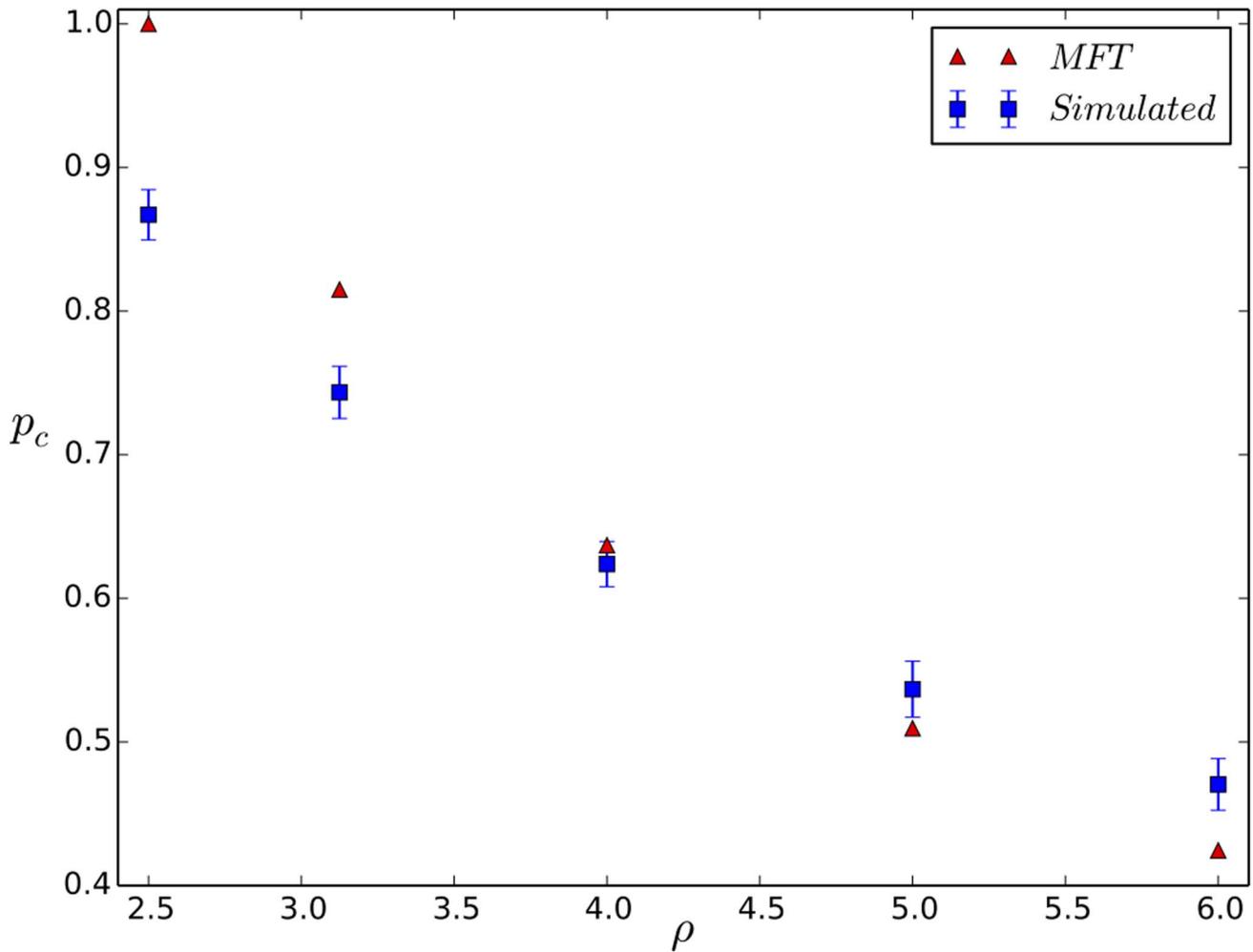

**Figure 9. Comparison of $p_c$ for Mean Field Theory with simulations.** We plot $p_c$ vs. $\rho$ for the MFT and the full simulations. Clearly, the MFT and full simulation trends are very similar.
doi:10.1371/journal.pone.0104965.g009

The collapse, if verified, has implications for treatment strategies for Alzheimer's. First, it is more likely to occur than MT dynamic instability, and thus represents the first irreversible signpost in late stage Alzheimer's. Second, treatments which aim to stabilize microtubules, such as those which are taxol based [32], will not inhibit this collapse. Rather, assuming as seems likely that phosphorylation is the source of tau removal from MTBs, approaches which inhibit activated kinases would appear more promising [33].

We plan future studies on combined mechanics/kinetics simulations of the MTBs under conditions of phosphorylation to attempt to determine the time scale for the collapse (which is not addressed here). We also will examine fully three dimensional models for the MTBs which include MT bending to provide a more realistic set of data for experimental tests on cultured cells.

## Materials and Methods

### 0.3 Percolation and Equilibration Algorithm

The percolation simulation algorithm is developed from an algorithm used by Tang and Thorpe in their simulation of rigidity percolation of an elastic triangular net [20]. Tang and Thorpe's simulation analyzed the percolation of a triangular net under a constant tension, as if stretched onto a frame. In this case, each node is treated as a lattice point attached to six stretched Hooke springs. In our case, however, each node is attached to six compressed Hooke springs, and subject to a distance-dependent attractive depletion force. Therefore, Tang and Thorpes algorithm requires some modification in order to be applicable to our case.

Our simulation begins by populating the center points of the nodes and the tau springs between them. As they are populated, an index is assigned to each node and each spring connection site. Then, a random number generator chooses a (random) spring index and removes one spring. This loss of a spring perturbs the balanced forces in the system and destroys the mechanical equilibrium. The simulation enters a loop, which runs over all node position indices. Beginning with $i = 0$, the simulation finds the net force vector acting on node $i$ by summing the force vectors from the spring forces, depletion forces, and steric forces, *viz.*

$$\vec{F}_i = \vec{F}_{spring,i} + \vec{F}_{D,i} + \vec{F}_{steric,i}. \qquad (9)$$

We then update the position $\vec{r}_i(t)$ to the new fictive time $t + dt$ by adding a distance proportional to the net force

$$\vec{r}_i(t + dt) = \vec{r}_i(t) + \alpha \vec{F}_i \qquad (10)$$

where $\alpha$ is an adjustable parameter used to control the step size. We use $\alpha = 10^{-6}$ nm/pN. This fictive molecular dynamics assumes the springs are overdamped or critically damped so that the nodes do not oscillate as they relocate.

Once the last node is moved, we check to see if the system has equilibrated to within a tolerance of $10^{-6}$ nm at each position during the previous loop; if not, we rerun the force relaxation loop until convergence at this tolerance is achieved. We average our results over 10–15 different starting random number seeds.

To estimate the second order elastic constant $C$ at a given $p$ value, we uniformly stretch or compress the area by a small dimensionless strain $\epsilon = \Delta a / a$ where $a$ is the average bond length, reequilibrate at the new strain value, and take $C$ from

$$C = \frac{[(E(+\epsilon) + E(-\epsilon)) - 2E(\epsilon = 0)]}{\epsilon^2} \qquad (11)$$

where $E$ is the total potential energy. We take $\epsilon$ sufficiently small that we are in the quadratic in strain region.

## Supporting Information

**Movie S1**
(MP4)

## Acknowledgments

We acknowledge useful conversations with S. Feinstein, N. Hall, A. Karsai, G.Y. Liu, C. DeCarli, and A. Levine.

## Author Contributions

Conceived and designed the experiments: AS DLC HRF RRPS NRH. Performed the experiments: AS HRF DLC. Analyzed the data: AS HRF DLC RRPS. Contributed reagents/materials/analysis tools: NRH RRPS. Contributed to the writing of the manuscript: DLC AS HRF RRPS NRH.

## References

1. Hardy JA, Higgins GA (1992) The amyloid cascade hypothesis. Science 256: 184–185.
2. Hardy JA, Selkoe DJ (2002) The amyloid hypothesis of alzheimer's disease: progress and problems on the road to therapeutics. Science 297: 353–356.
3. Fein JA, Sokolow S, Miller CA, Vinters HV, Yang F, et al. (2008) Co-localization of amyloid beta and tau pathology in alzheimer's disease synaptosomes. Am J Pathology 172: 1683–1692.
4. McKee AC, Cantu RC, Nowinski CJ, Hedley-Whyte ET, Gavett BE, et al. (2009) Chronic traumatic encephalopathy in athletes: Progressive tauopathy following repetitive head injury. J Neuropathol Exp Neurol 68: 709–735.
5. Wang JZ, Liu F (2008) Microtubule-associated protein tau in development, degeneration, and protection of neurons. Progress in Neurobiology 85: 148–175.
6. Reifert J, Hartung-Cranston D, Feinstein SC (2011) Amyloid $\beta$-mediated cell death of cultured hippocampal neurons reveal extensive tau fragmentation without increased full-length tau phos-phorylation. J Biol Chem 286: 20792–20811.
7. Zempel H, Thies E, Mandelkow E, Mandelkow EM (2010) A-beta oligomers cause localized ca$^{2+}$ elevation, missorting of endogenous tau into dendrites, tau phosphorylation, and destruction of microtubules and spines. J Neurosci 30: 11938–11950.
8. Hirokawa N (1994) Microtubule organization and dynamics dependent on microtubule-associated proteins. Curr Op Cell Biol 6: 74–81.
9. Pontes B, Ayala Y, Fonseca ACC, Romao LF, Amaral RF, et al. (2013) Membrane elastic properties and cell function. PLoS One 8: e67708.
10. Press W, Teukolsky S, Vetterling W, Flannery B (2007) Numerical Recipes. Cambridge: Cambridge Univ. Press, 3rd edition, 516–520 pp.
11. Brady ST, Colman DR, Brophy PJ (2013) Subcellular organization of the nervous system: Organelles and their functions. In: Squire L, editor, Fundamental Neuroscience, Burlington MA: Academic Press. 3rd edition, p. 97.
12. Rosenberg KJ, Ross JL, Feinstein HE, Feinstein SC, Israelachvili J (2008) Complementary dimerization of microtubule-associated tau protein: Implications for microtubule bundling and tau-mediated pathogenesis. Proc Natl Acad Sci USA 105: 7445–7450.
13. Phillips R, Kondev J, Theriot J (2008) Physical Biology of the Cell. New York: Garland Science, 1st edition, 313–315 pp.
14. Dietz H, Berkemeier F, Bertz M, Rief M (2006) Anisotropic deformation response of single protein molecules. Proc Nat Acad Sci USA 34: 12724–12728.
15. Chtcheglova LA, Shubeita GT, ASekatskii S, Dietler G (2004) Force spectroscopy with a small dithering of afm tip: A method of direct and continuous measurement of the spring constant of single molecules and molecular complexes. Biophys J 86: 1177–1184.
16. Phillips R, Kondev J, Theriot J (2008) Physical Biology of the Cell. New York: Garland Science, 1st edition, 526–528 pp.
17. Mylonas E, Hascher A, Bernado P, Blackledge M, Mandelkow E, et al. (2008) Domain conformation of tau protein studied by solution small-angle x-ray scattering. Biochemistry 47: 10345–10353.
18. King ME, Ahuja V, Binder LI, Kuret J (1999) Ligand-dependent tau filament formation: implications for alzheimer's disease progression. Biochemistry 38: 14851–14859.
19. Needleman D, Ojeda-Lopez MA, Raviv U, Ewert K, Miller HP, et al. (2005) Radial compression of microtubules and the mechanism of action of taxol and associated proteins. Biophys J 89: 3410–3423.
20. Tang W, Thorpe MF (1988) Percolation of elastic networks under tension. Phys Rev B 37: 5539–5551.
21. Chau MR, Radeke MJ, deInes C, Barasoain I, Kohlstaedt LA, et al. (1998) The microtubule-associated protein tau crosslinks to two distinct sites on each $\alpha$ and $\beta$ tubulin monomer via separate domains. Biochemistry 37: 17692–17703.
22. Kar S, Fan J, Smith MJ, Goedert M, Amos LA (2003) Repeat motifs of tau bind to the insides of microtubules in the absence of taxol. EMBO J 22: 70–77.
23. Hawkins TL, Sept D, Mogessie B, Straube A, Ross J (2013) Mechanical properties of doubly stabilized microtubule filaments. Biophys J 104: 1517–1528.
24. Smith J, Wilson L, Azarenko O, Zhu XJ, Lewis BM, et al. (2010) Eribulin binds at microtubule ends to a single site on tubulin to suppress dynamic instability. Biochemistry 49: 1331–1337.
25. Sykes MF, Essam JW (1964) Exact critical percolation probabilities for site and bond problems in two dimensions. J Math Phys 5: 1117–1127.
26. Bero AW, Yan P, Roh JH, Cirrito JR, Steward FR, et al. (2011) Neuronal activity regulates the regional response to amyloid-beta deposition. Nature Neurosci 14: 750–756.
27. Jin M, Shepardson N, Yang T, Chen G, Walsh D, et al. (2011) Soluble amyloid beta-protein dimers isolated from alzheimer cortex directly induce tau hyperphosphorylation and neuritic damage. Proc Nat Acad Sci (USA) 108: 5819–5824.
28. Lulevich V, Zimmer CC, Hong HS, Jin LW, Liu GY (2010) Single-cell mechanics provides a sensitive and quantitative means for probing amyloid-$\beta$ peptide and neuronal cell interactions. Proc Nat Acad Sci USA 107: 13872–13877.
29. Huang J, Friedland RP, Auchus AP (2007) Diffusion tensor imaging of normal-appearing white matter in mild cognitive impairment and early alzheimer disease: Preliminary evidence of axonal degeneration in the temporal lobe. Am J Neuroradiology 28: 1943–1948.
30. Salat DH, Greve DN, Pacheco JL, Quinn BT, Helmer KG, et al. (2009) Regional white matter volume differences in nondemented aging and alzheimer's disease. NeuroImage 44: 1247–1258.
31. Lingor P, Koch JC, Toenges L, Baehr M (2012) Axonal degeneration as a therapeutic target in the cns. Cell Tissue Res 349: 289–311.
32. Michaelis ML (2006) Ongoing *in vivo* studies with cytoskeletal drugs in transgenic mice. Current Alzheimer Research 3: 215–219.
33. Mueller BK, Mack H, Teusch N (2005) Rho kinase, a promising drug target for neurological disorders. Nature Rev Drug Discov 4: 387–398.